\documentclass[reprint,superscriptaddress,amsmath,amssymb,prl,longbibliography,nofootinbib]{revtex4-1}

 \usepackage[bottom]{footmisc}

\usepackage[T1]{fontenc}
\usepackage[latin9]{inputenc}
\usepackage{mathrsfs}
\usepackage{bm}
\usepackage{amsmath}
\usepackage{amsthm}
\usepackage{amssymb}
\usepackage{stmaryrd}
\usepackage{mathtools}
\usepackage{dsfont}

\edef\ordinarycolon{\mathchar\the\mathcode`: }
\edef\ordinaryequals{\mathchar\the\mathcode`= }

\usepackage{breqn}
\catcode`^=7

\AtBeginDocument{%
  \catcode`^=12
}

\makeatletter

\allowdisplaybreaks

\let\cat@comma@active\@empty

\usepackage[capitalize]{cleveref}
\usepackage{color}
\RequirePackage[dvipsnames,usenames]{xcolor}

\newif\ifnotes
\notestrue

\makeatother

\newcommand{\ba}{\begin{eqnarray}}
\newcommand{\ea}{\end{eqnarray}}

\newcommand{\eq}[1]{\begin{align}#1\end{align}}

\newtheorem{theorem}{Theorem}

\begin{document}

\preprint{}

\title{Stochastic thermodynamics and fluctuation theorems for non-linear systems}

\author{Jan Korbel}
 \affiliation{Section for Science of Complex Systems, CeMSIIS, Medical University of Vienna, Spitalgasse 23, 1090 Vienna, Austria}
\email{jan.korbel@meduniwien.ac.at}
\affiliation{Science Hub Vienna, Josefst\"{a}dter Strasse 39, 1080 Vienna, Austria}
\author{David H. Wolpert}%
\affiliation{Santa Fe Institute, Santa Fe, NM, United States of America}
\email{david.h.wolpert@gmail.com}
\homepage{ http://davidwolpert.weebly.com}
\affiliation{Science Hub Vienna, Josefst\"{a}dter Strasse 39, 1080 Vienna, Austria}
\affiliation{ Arizona State University, Tempe, AZ, United States of America}






\date{\today}

\begin{abstract}
We extend stochastic thermodynamics by relaxing the two assumptions that the Markovian dynamics must be linear
and that the equilibrium distribution must be a Boltzmann distribution. We show that if we require the second law to hold
when those assumptions are relaxed, then it cannot be formulated in terms of Shannon
entropy. However, thermodynamic consistency is salvaged if we reformulate the second law in terms of generalized entropy; our first result is
an equation relating the precise form of the non-linear master equation to the precise associated generalized entropy which results in
thermodynamic consistency. We then build on this
result to extend the usual trajectory-level definitions of thermodynamic quantities that are appropriate even when the two assumptions
are relaxed. We end by using these trajectory-level definitions to derive extended
versions of the Crooks fluctuation theorem and Jarzynski equality which apply when the two assumptions
are relaxed.

\end{abstract}

\maketitle











Stochastic thermodynamics has uncovered many important results concerning non-equilibrium systems, including various fluctuation theorems \cite{Evans94,Crooks99,Jarzynski00,Evans02,Seifert05}, thermodynamic uncertainty relations \cite{Barato15,Horowitz20,Falasco20,Ito20}, and speed limits \cite{Shiraishi18,Shiraishi19}.
These results have been applied in a broad range of disciplines, including physics \cite{Ciliberto17}, chemistry \cite{Schmiedl07}, biophysics \cite{Seifert12}, active matter \cite{Speck16}, quantum thermodynamics \cite{Strasberg19} and computation \cite{Wolpert19,Wolpert20}.

However, there are several assumptions commonly made in stochastic thermodynamics, which restricts its applicability.
In this paper we consider relaxing two of those assumptions, to expand the range of scenarios
that can be analyzed using stochastic thermodynamics. However, as we prove, in those
scenarios the second law would be violated  --- if the second law were formulated in terms of conventional
Shannon entropy. As we show, thermodynamic consistency can be maintained, but only if
we use a suitable generalized entropy. In addition, we show that appropriately extended
forms of the Jarzynski equality and Crook's theorem still hold in this broader setting
if use such a generalized entropy.


Most of the research on stochastic thermodynamics to date has assumed that the system of interest is connected to one or more heat baths that are much larger than that system of interest (e.g., ``infinite'' heat baths) \cite{Broek13}. Microreversibility, i.e., ``local detailed balance'', then requires that the equilibrium
distributions associated with each of those baths is a Boltzmann distribution \cite{Seifert08}. However,
previous research motivates the choice of non-Boltzmann local equilibrium distributions when considering
systems at equilibrium, e.g., in connection with finite heat bath
 thermodynamics \cite{Adib03,Campisi09,Richens18}, long-range interactions \cite{Jiulin07,Biro13,Jizba19} or generalized thermostatistics \cite{Almeida99,Rama00,Shiino98,Campisi07}.
To date, none of this research has been extended to open, non-equilibrium systems, the domain of interest in this paper.
(See~\cite{campisi_finite_2009} for a partial exception, where the starting distribution is defined by a finite heat bath but the subsequent dynamics is closed.
)
Accordingly, the first assumption we relax is that local detailed balance is defined in terms of a Boltzmann distribution, either due
to the infinite size of the baths or (if they are finite) for some other reason.

Another typical assumption in stochastic thermodynamics has been that the probability distribution evolves according to a \textit{linear} master equation. However, this assumption is not always warranted. Accordingly,
several authors have explored non-linear \cite{Ottinger10,Smadbeck14,Sanchez15} or non-Markovian \cite{Esposito03,Strasberg16,Whitney18} thermodynamics. There has also been some very restricted earlier research
investigating the non-linear master equation \cite{Frank04}. In addition,
\cite{Peng20} analyzed stochastic thermodynamics with non-linear internal energy and non-extensive Tsallis entropy \cite{Tsallis88},
obtaining the $q$-deformed exponential equilibrium distribution. \cite{Curado03} considered a master equation with probability-dependent transition rates and showed that this is equivalent to a certain class of non-linear processes.
These special cases notwithstanding, there has been very little research that systematically
investigating stochastic thermodynamics of non-linear systems.
Accordingly, the second assumption we relax is to allow the Markovian dynamics to be nonlinear.

Given these two relaxations of conventional assumptions, in order to impose the second law in a thermodynamically consistent way, we need
to consider other entropies besides Shannon entropy. There are many different possible choices for such a generalized class of entropies \cite{Burg72,Renyi76,Havrda76,Sharma78,Tsallis88,Kaniadakis02,Frank00,Jizba04,Tsallis05,Liese06,Jizba16,Hanel2011a,Korbel18,Thurnerbook,Tempesta20}.
However, it was recently shown that the class of entropies that fulfills
the first three Shore-Johnson axioms \cite{Shore80,Shore81} (equivalent to a generalization of Shannon-Khinchin axioms \cite{Jizba20}) can be written in the form $\mathbb{S} = f(\sum_i g(p_i))$, which is known as the \emph{sum-class of entropies}
 \cite{Jizba19}. For these reasons, in order to ensure that the second law is satisfied despite our two relaxed assumptions, we generalize beyond Shannon entropy to arbitrary sum-class entropies.

The rest of this letter is organized as follows. First, we formulate the three requirements that a thermodynamic system should fulfill, i.e., we introduce our generalized definitions of Markovian dynamics, local detailed balance, and the second law. Second, we derive the relation between the form of entropic functional and the form of non-linear master equation. Finally, we introduce stochastic versions of thermodynamic quantities and derive fluctuation theorems for non-linear systems.

\textit {Thermodynamic requirements. ---}
We consider a system with a finite
number of states, labeled $m, n$, etc.  All quantities are implicitly functions of time $t$, and for any variable $z$ which varies with $t$,
we use $\dot{z}$ to indicate $\mathrm{d} z(t) /\mathrm{d}t$.
Before formulating our requirements, we define several terms,
all based on an arbitrary time-dependent scalar-valued \textit{internal energy} function of the states, $\epsilon_{m}$.
At this point, despite the terminology, there is no physical meaning associated with $\epsilon_{m}$;
such meaning will arise through Requirements 2 and 3 below.

We write the expected internal energy as
\eq{
\label{eq:2}
U &:= \sum_m p_m \epsilon_m\,
}
and we define $\dot{Q} := \sum_m \dot{p}_m \epsilon_m$ and $\dot{W} := \sum_m p_m \dot{\epsilon}_m$ as the \textit{rate of work} and as the \textit{heat flow}, respectively\footnote{Some authors have considered different forms for the internal energy , e.g., by using escort distributions \cite{Tsallis98,Martinez00}, generalized means \cite{Czachor02,Scarfone16} or tail constraints \cite{Bercher08}. In our analysis, we stick to the common definition of internal energy because of its clear physical interpretation.}.
So the ``first law of thermodynamics'' holds $\dot{U} = \dot{Q} + \dot{W}$.

Next, we define the (generalized) \textit{information entropy} as
    \begin{equation}
    \mathbb{S} := f\left(\sum_m g(p_m)\right)
\label{eq:entropy}
\end{equation}
which reduces to ordinary Shannon information entropy $\mathbb{H} = - \sum_m p_m \log p_m$,
for $f(x) = x$ and $ g(x) = - x \log x$. The form of entropy used in Eq.~\eqref{eq:entropy} is known as \emph{sum-class} \cite{Tempesta11}, a generalization of \emph{trace-class} \cite{Hanel2011a}, which is the special case where $f(x) = x$.
Maximizing the generalized entropy subject to a given value of the expected internal energy
gives us the \textit{equilibrium} distribution:
    \begin{equation}
        \pi_m = (g')^{-1} \left(\frac{\alpha+\beta \epsilon_m}{C_f(\pi)}\right)
\end{equation}
where $C_f(\pi) = f'\left(\sum_m g(\pi_m)\right)$.
In order to obtain a unique equilibrium distribution, we also impose the following requirements on functions $f$ and $g$:
\begin{itemize}
    \item $f$ is a strictly increasing function, i.e., $f' >0$,
    \item $g$ is a concave function, i.e.,  $g'' \leq 0$.
\end{itemize}

Next, we introduce requirements concerning the thermodynamics of a system coupled to one or more heat baths.
These requirements provide us with the promised physical interpretation of the ``internal energy'', ``work'', and ``heat flow''.\\
\\
\noindent \textbf{Requirement 1:} \emph{Markovian evolution.}
We assume that the dynamics of our system evolves according to a first-order differential equation, as in conventional stochastic thermodynamics, but allow that equation to be nonlinear:
    \begin{equation}
        \dot{p}_m = \sum_n [w_{mn}\Omega(p_n) - w_{nm}\Omega(p_m)]
\label{eq:1}
\end{equation}
where $\Omega(p_n)$ is an arbitrary function while $w_{mn}$ is a set of real-valued parameters. {In the Supplemental material (SM) \cite{SM}, we show how this non-linear master equation can be derived from Markov chains with probability-dependent transition probabilities.} Overview of physical and life-science applications of Markov chains with probability-dependent transition rates can be found e.g., in review \cite{Frank13}.

Note that the antisymmetric form of the summand in \cref{eq:1} ensures probability conservation: regardless of the precise
form of $\Omega$, $\sum_n \dot{p}_n =0$. We refer to each $w_{mn}$ as a (generalized) \textit{transition rate}. In
general $w_{mn}$ will not have the form of a conventional transition rate.  (The precise connection between the matrix $w_{mn}$ and
conventional transitions rates is discussed in the SM.)\\
\\
\noindent \textbf{Requirement 2:} \emph{Local Detailed Balance.}
We assume that the system is coupled to a finite set of infinite heat reservoirs, $\mathcal{V}$,
with associated temperatures $T^{\nu}$, $\nu \in \mathcal{V}$. We formalize this assumption in several steps. First, we assume the transition rates are given
by a sum over rates for each reservoir, i.e., $w_{mn} = \sum_\nu w^\nu_{mn}$. Next, define the  \textit{local equilibrium} of the system for each such reservoir $\nu$ as the distribution that would be the equilibrium of the system
if it were only connected to that reservoir, i.e.,
as the distribution $p_m$ with maximal entropy (as defined in \cref{eq:entropy}), subject to the constraint that the expected energy defined in \cref{eq:2} equals $\beta^\nu = 1 / T^\nu$ (up to
a reservoir-independent proportionality constant).

In the usual way we can solve for that maximizing distribution, getting $\pi^{\nu}_m = (g')^{-1} \left((\alpha^\nu+\beta^\nu \epsilon_m)/(C_f^\nu(\pi^\nu)\right)$ where $C_f^\nu(\pi^\nu) = f'\left(\sum_m g(\pi^{\nu}_m)\right)$. The requirement of local detailed balance means that the local probability flow $J_{mn}^\nu = w^\nu_{mn}\Omega(p_n) - w^\nu_{nm}\Omega(p_m)$,
vanishes for the local equilibrium distribution, i.e.,
    \begin{equation}\label{eq:ldb}
        \frac{w_{mn}^\nu}{w^{\nu}_{nm}} = \frac{\Omega(\pi^{\nu}_m)}{\Omega(\pi^{\nu}_n)}\, .
    \end{equation}

\noindent \textbf{Requirement 3:} \emph{Second law of thermodynamics.}
We introduce a few more definitions in order to formalize the second law. We define the \emph{heat flow from a reservoir $\nu$} as
\eq{
\dot{Q}^\nu := \sum_{mn} [w_{mn}^\nu\Omega(p_n) - w_{nm}^\nu\Omega(p_m)] \epsilon_m
}
so that the total heat flow is $\dot{Q} = \sum_{\nu} \dot{Q}^\nu$.
Similarly, we define the \emph{entropy flow from reservoir $\nu$} into the system as $\dot{S}_e^\nu  := \frac{\dot{Q^{\nu}}}{T^{\nu}}$
and the overall \textit{entropy flow} as $\dot{S}_e = \sum_\nu \dot{S}_e^\nu =  \sum_{\nu}\frac{\dot{Q^{\nu}}}{T^{\nu}}\,$.
 We then define the \textit{entropy production} as
 \begin{equation}\label{eq:ep}
\dot{S}_i := \frac{\mathrm{d} \mathbb{S}}{\mathrm{d} t} - \dot{S}_e.
 \end{equation}
where the entropy has the form given in \cref{eq:entropy}. Using these definitions,
the generalized second law is the requirement that
\begin{eqnarray}
\dot{S}_i &\geq& 0,\nonumber\\
\dot{S}_i &=& 0 \iff \ J_{mn}^\nu = 0 \quad \forall \nu,n,m.
    \end{eqnarray}

\textit{Necessary relationship between nonlinear master equation and entropy. ---}
Our first main result is a relation between the function $\Omega$ specifying the nonlinear Markovian dynamics
and the precise generalized entropy which must hold if the three requirements introduced
above all hold:
\begin{theorem}
Requirements 1-3 imply that
\begin{equation}
   \Omega(p_m) = \exp(-g'(p_m))\, .
\label{eq:main}
\end{equation}
\label{thm:main}
\end{theorem}
\noindent \cref{thm:main} means that, in general, a given nonlinear Markovian dynamics will
require that we not use Shannon entropy, if we wish to maintain thermodynamic consistent. \cref{thm:main}
is also the basis for our other results below. (Note that \cref{thm:main} only relates $\Omega$ to $g$; $f$ remains an arbitrary increasing function.)

The complete proof of \cref{thm:main} is in the SM \cite{SM}. Here we outline its main steps. First, we calculate the time derivative of generalized information entropy. Second, we decompose that derivative into two terms, which as we show equal the (non-negative) entropy production rate and the entropy flow rate. (The requirement of local detailed balance is invoked to identify the entropy flow rate with the heat rate over the temperature.) From this we deduce the form of the function $\Omega$ presented in \cref{eq:1}.

A corollary of  \cref{thm:main} arises if we plug it into \cref{eq:1}, to derive the general form that the master equation must have in order to meet the three requirements:
\begin{equation}
        \dot{p}_m = \sum_{n,\nu} \left[w^\nu_{mn} \exp(-g'(p_n)) - w^\nu_{nm} \exp(-g'(p_m))\right]\, .
\label{eq:non-lin}
\end{equation}
There are multiple examples in the literature of
systems that can be described by a master equation \eqref{eq:non-lin} or corresponding Fokker-Planck equation. (See SM for derivation of that Fokker-Planck equation \cite{SM}.) These include chemical reaction networks \cite{Zhou16}, morphogen gradient formation \cite{Boon12}, financial markets and option pricing \cite{Borland98} ($g(p_i) \propto p_i^q$); heavy-ion collisions \cite{Bertsch88}, turbulence and vortex formation \cite{Jin98,Miller90,Chavanis03} ($g(p_i) = p_i \log(p_i) + (1-p_i) \log(1 - p_i)$); or negative-feedback systems \cite{Frank02} ($g(p_i) = p_i \log(p_i/(1+\alpha p_i)$).

Another corollary of \cref{thm:main} is that if we know the form of $\Omega$, then the entropic functional satisfying Requirements 1-3 can be written in the following form:
\begin{equation}
\mathbb{S} = f\left(-\sum_m \int_0^{p_m} \mathrm{d} z\, \log  \Omega(z)\right)\, .
\end{equation}

Note also by plugging \cref{thm:main} into the requirement of local detailed balance that local detailed balance
has the same form as in conventional stochastic thermodynamics:
\begin{equation}
\frac{w_{mn}^\nu}{w_{nm}^\nu} = \exp\left(- \frac{\epsilon_m-\epsilon_n}{T^\nu}\right)\, .
\end{equation}
In addition, we can express the entropy production rate as
\begin{equation}
\dot{S}_i = \frac{C_f(p)}{2} \sum_{mn\nu} (w_{nm}^\nu \Omega(p_m) - w_{mn}^\nu \Omega(p_n)) \log\left(\frac{w_{nm}^\nu \Omega(p_m)}{w_{mn}^\nu \Omega(p_n)} \right)
\end{equation}
and the entropy flow rate as
\begin{equation}
\dot{S}_e = \frac{C_f(p)}{2} \sum_{mn\nu} (w_{nm}^\nu \Omega(p_m) - w_{mn}^\nu \Omega(p_n)) \log\left(\frac{w_{nm}^\nu }{w_{mn}^\nu} \right)
\end{equation}
where $C_f(p) = f'(\sum_m g(p_m))$.

\textit{Stochastic thermodynamics for nonlinear master equations. ---}
We now introduce the stochastic (trajectory-level) versions of the ensemble thermodynamic quantities
investigated above. For the rest of the paper, we assume $f(x) = x$ (i.e., the trace-class entropy), for simplicity.

To begin, we define the \textit{stochastic entropy} of a stochastic trajectory $n(.)$ at time $\tau \in [0,T]$
as
\begin{equation}
s(n(\tau)) := \log \left( \frac{1}{\Omega(p_{n(\tau)}(\tau))}    \right) = g'(p_{n(\tau)}(\tau))
\end{equation}
and stochastic energy as $e(n(\tau)) := \epsilon_{n(\tau)}$.
The stochastic heat flow and work flow can be obtained from the first law on the trajectory level $\dot{e}(\tau) = \dot{q}(\tau) + \dot{w}(\tau)$ where $\dot{q}(\tau) = \sum_m \dot{\delta}_{m,n(\tau)} \epsilon_{m}$ and $\dot{w}(\tau) = \sum_m \delta_{m,n(\tau)} \dot{\epsilon}_{m}\,$, respectively. Thus, the ensemble first law of thermodynamics can be written as $\langle \dot{e} \rangle = \langle \dot{q} \rangle + \langle \dot{w} \rangle$ where $\langle \cdot \rangle$ denotes the ensemble average, i.e., multiplication of $p_{n(\tau)}$ and summing over all trajectories $n(\tau)$.

The average energy can be obtained as the ensemble average of the stochastic energy, i.e., $U(\tau) = \langle e(\tau) \rangle$. However, in general such equivalence of averages does not hold for entropy, once one goes beyond the special case of Shannon information entropy, i.e., in general $\mathbb{S}(\tau) \neq \langle s(n(\tau)) \rangle\,$ . Nevertheless, the second law of thermodynamics is still valid on the trajectory level. In the SM \cite{SM} we show that $\dot{s}(n(\tau))=\dot{s}_i(n(\tau))+\dot{s}_e(n(\tau))$  where $\dot{s}_i(n(\tau))$ is the entropy production rate and  $\dot{s}_e(n(\tau))$. Due to the local detailed balance, we obtain that $\dot{s}_e(n(\tau)) = \sum_\nu \beta_\nu \dot{q}^\nu(n(\tau))$. The ensemble second law \eqref{eq:ep} can be obtained by averaging over all possible trajectories.

Now consider an external protocol $\lambda(\tau)$ which drives the energy spectrum, i.e., $\epsilon_m(\tau) \equiv \epsilon_m(\lambda(\tau))$. In the usual way define a time-reversed trajectory $\tilde{n}(\tau) = n(T-\tau)$ and the associated time-reversed driving protocol $\tilde{\lambda}(\tau) = \lambda(T-\tau)$.
Suppose that local detailed balance is fulfilled. Then as we show in the SM \cite{SM},
\begin{equation}\label{eq:dft}
\log \frac{P(n)}{\tilde{P}(\tilde{n})} = \Delta s_i
\end{equation}

where $\tilde{P}$ denotes probability under time-reversed protocol $\tilde{\lambda}(\tau)$. This means that the standard form of the detailed fluctuation theorem \cite{Esposito10b}
must hold, just with our modified definition of entropy production:
\begin{equation}\label{eq:dft}
\frac{P(\Delta s_i)}{\tilde{P}(-\Delta s_i)} = e^{\Delta s_i}\, .
\end{equation}

Next, write the trajectory version of the Massieu function (also called  \emph{free entropy} \cite{Naudts11book}) as $\psi = s - \beta e$. The entropy production can be written as
\begin{equation}\label{eq:si}
\Delta s_i = \beta w + \Delta \psi\, .
\end{equation}
Consider the situation where the initial distribution is equilibrium one.
For equilibrium state the stochastic Massieu function is independent of the state since it is equal the Lagrange parameter $\alpha$, i.e., $\psi(\pi_m) = s(\pi_m) - \beta \epsilon_{m} = \alpha = \langle \psi \rangle$. Combining
this with  \cref{eq:dft} and \cref{eq:si} gives (a generalized form of) the Crooks fluctuation theorem,
\begin{equation}
\frac{P(w)}{\tilde{P}(-w)} = \exp(\beta w + \Delta \alpha)
\end{equation}
As usual this in turn gives (a generalized form of) the Jarzynski equality
\begin{equation}
\langle \exp(-\beta w) \rangle = \exp(\Delta \alpha).
\end{equation}
Jensen's inequality then gives
\begin{equation}
W = \langle w \rangle \geq - T \Delta \alpha\, .
\end{equation}
These results provide a clear physical interpretation of the Lagrange parameter $\alpha$: for a process that starts and ends in an equilibrium, $-T\Delta \alpha$ is the minimal amount of work necessary for transition between two equilibrium states.

We finally mention that the validity of DFT \eqref{eq:dft} has further consequences. Consider the situation when initial and final distribution agree $p_n(0) = p_n(T)$ and when the external protocol is time-symmetric, i.e., ${l}(\tau) = \tilde{l}(\tau)$. Let us consider a trajectory quantity that is antisymmetric under time-reversal, i.e., $\phi(n(\tau)) = - \phi(\tilde{n}(\tau))$. Then the Fluctuation theorem uncertainty relation holds:
\begin{equation}
\frac{\mathrm{Var}(\phi)}{\langle \phi \rangle^2} \geq \frac{2}{e^{\langle \Delta s_i\rangle}-1}
\end{equation}
as shown in \cite{Hasegawa19}. In a similar spirit, dissipation asymmetry relation \cite{Campisi20}, time-dissipation uncertainty relation \cite{Falasco20a} or the speed-limit theorem \cite{Shiraishi18} can be also straightforwardly extended to systems driven by non-linear master equation \eqref{eq:1}, see the SM \cite{SM}.

\textit{Discussion. ---} In this paper we considered a broad type of nonlinear Markovian dynamics. We also define a generalized ``local equilibrium
distribution'', as the distribution
that maximizes the ``information entropy'', which we allow to be any specified sum-class entropy functional.
This allowed us to impose a generalized form of local detailed balance, by requiring that the currents for
each reservoir vanish if the system is in its local equilibrium distribution for the specified information entropy.
We required that the generalized, nonlinear Markovian dynamics obey this generalized form of local detailed balance.
Next, we define
``entropy flow''  in the usual way in stochastic thermodynamics, in terms of heat flow into the reservoirs. Given these definitions, we define the (generalized)
second law by requiring that under the nonlinear Markovian dynamics, the time-derivative of the difference between the specified
information entropy and the entropy flow is non-negative.

Our first result was to show that for this generalized form of the second law to hold, there must be a certain simple relationship between the choices of which particular sum-class
entropy to use to define the information
entropy, and of the particular nonlinear form of the Markovian dynamics. We then used this result to derive trajectory-level definitions of (generalized) thermodynamic quantities.
Our second result is that those generalized stochastic thermodynamic quantities obey both the Jarzynski equality and Crooks' theorem.

It is important to emphasize that although we broadened the range of scenarios that stochastic thermodynamics can apply to,
we still restricted those scenarios in several ways. For example
\cref{eq:1} is not the most general form of a non-linear master equation. Other types of non-linear equations are obtained from mean-field approximations of many-body systems, phase transitions and self-organization or approximations of quantum master equations, see \cite{Frank04book} for an overview. The question at stake is whether the results can be extended to more general types of non-linear Markovian and non-Markovian systems. This might require to consider entropic functionals of more generalized form, or even generalizations of the maximum entropy principle to e.g., principle of maximum caliber \cite{Dixit17}.

\begin{acknowledgments}
\textit{Acknowledgements. ---}
DHW thanks the Santa Fe Institute for support and acknowledges support from the U.S. National Science Foundation, Grant No. CHE-1648973  and from the FQXi foundation, Grant No. FQXi-RFP-1622. JK was supported by the Austrian Science Fund (FWF) under project I3073. The opinions expressed in this paper are those of the authors and
do not necessarily reflect the view of the National Science Foundation.
\end{acknowledgments}

\onecolumngrid
\newpage
\appendix

\section{Non-linear Markov chains and non-linear Master equation}

In this appendix we show how to derive a non-linear master equation from a Markov chain with probability-dependent transition rates. More details about non-linear Markov chains and their expression in terms of stochastic processes with probability-dependent transition rates can be found e.g., in \cite{Frank13}. Consider a discrete-time stochastic process whose $n$-point distribution can be written in a form of non-linear Markov process, i.e.,
\begin{equation}\label{eq:nlmc}
p(x_n,t_n;x_{n-1},t_{n-1};\dots;x_0,t_0) = \prod_{j=1}^n W(x_j,t_j|x_{j-1},t_{j-1}) \Omega(p(x_0,t_0))
\end{equation}
Such a process is called ``non-linear Markov chain''. 
Note that the function $W(x_j,t_j|x_{j-1},t_j)$ cannot be interpreted as transition probability. This can be easily shown e.g., for $n=2$ and summing \eqref{eq:nlmc} over $x_1$. We get that
\begin{equation}
\sum_{x_1} W(x_1,t_1|x_{0},t_{0}) = \frac{p(x_0,t_0)}{\Omega(p(x_0,t_0))}.
\end{equation}

Now consider a two-point distribution
\begin{equation}
  p(x_1,t_1;x_0,t_0) =  W(x_1,t_1|x_{0},t_0) \Omega(p(x_0,t_0))
\end{equation}
Then, the transition probability $T$ that is defined in the usual way can be expressed as follows:
\begin{equation}
T(x_1,t_1|x_{0},t_0) =  \frac{p(x_1,t_1;x_0,t_0)}{p(x_0,t_0)} = W(x_1,t_1|x_{0},t_0) \omega(p(x_0,t_0))
\end{equation}
where $\omega(z) = \Omega(z)/z$. Thus, the transition probability explicitly depends on the probability distribution $p(x_0,t_0)$, i.e.,
\begin{equation}
T(x_1,t_1|x_{0},t_0) \equiv T(x_1,t_1|x_{0},t_0,p(x_0,t_0))\, .
\end{equation}

Next consider a three point distribution
\begin{equation}
p(x_2,t_2;x_1,t_1;x_0,t_0) = W(x_2,t_2|x_{1},t_1) W(x_1,t_1|x_{0},t_0) \Omega(p(x_0,t_0))\, .
\end{equation}
This can be rewritten in terms of transition probabilities as
\begin{equation}
 p(x_2,t_2;x_1,t_1;x_0,t_0) = \frac{T(x_2,t_2|x_{1},t_1,p(x_1,t_1)) T(x_1,t_1|x_{0},t_0,p(x_0,t_0))}{\omega(p(x_1;t_1))}p(x_0,t_0)\, .
\end{equation}
By summation over $x_1$, 
we obtain the Chapman-Kolmogorov equation for non-linear Markov chains:
\begin{equation}
T(x_2,t_2|x_{0},t_0,p(x_0,t_0)) = \sum_{x_1} \frac{T(x_2,t_2|x_{1},t_1,p(x_1,t_1)) T(x_1,t_1|x_{0},t_0,p(x_0,t_0))}{\omega(p(x_1;t_1))}
\end{equation}
%
From the self-consistency requirement that
\begin{equation}
p(x_1,t_1) = \lim_{t_2 \rightarrow t_1} T(x_2,t_2|x_{1},t_1,p(x_1,t_1)) p(x_1,t_1)
\end{equation}
we obtain that $\lim_{t_2 \rightarrow t_1} T(x_2,t_2|x_{1},t_1,p(x_1,t_1)) = \delta(x_2-x_1)$.

For small times, we can approximate the transition probability by the first-order expansion in terms of $t_2$ around $t_1$
\begin{equation}
T(x_2,t_2|x_{1},t_1,p(x_1,t_1)) =  (t_2-t_1) \tilde{T}(x_2|x_1,t_1,p(x_1,t_1)) + \mathcal{O}((t_2-t_1)^2)
\end{equation}
By adding a zero term to the right-hand side, we get
\begin{equation}
T(x_2,t_2|x_{1},t_1,p(x_1,t_1)) \approx (t_2-t_1) \tilde{T}(x_2|x_1,t_1,p(x_1,t_1)) +  \left(1-(t_2-t_1)\sum_{x'} \tilde{T}(x'|x_1,t_1,p(x_1,t_1))\right) \delta(x_2-x_1).
\end{equation}
By plugging this into the Chapman-Kolmogorov equation, we obtain
\begin{eqnarray}
T(x_2,t_2|x_{0},t_0,p(x_0,t_0)) &\approx& (t_2-t_1) \sum_{x_1} \left( \frac{\tilde{T}(x_2|x_1,t_1,p(x_1,t_1))T(x_1,t_1|x_{0},t_0,p(x_0,t_0))}{\omega(p(x_1,t_1))} \right.\nonumber\\
&-& \left.\frac{\tilde{T}(x_1|x_2,t_2,p(x_2,t_2))T(x_2,t_2|x_{0},t_0,p(x_0,t_0))}{\omega(p(x_2,t_2))}  \right) \;+\;   T(x_2,t_1|x_{0},t_0,p(x_0,t_0))
\end{eqnarray}
Reorder the terms in this equation and take the limit, to establish that 
\begin{eqnarray}
&\lim_{t_2\rightarrow t_1} \frac{\tilde{T}(x_2,t_2|x_{0},t_0,p(x_0,t_0))-T(x_2,t_1|x_{0},t_0,p(x_0,t_0))}{t_2 - t_1}  \nonumber \\
&= \sum_{x_1} \left( \frac{T(x_2|x_1,t_1,p(x_1,t_1))T(x_1,t_1|x_{0},t_0,p(x_0,t_0))}{\omega(p(x_1,t_1))} - \frac{\tilde{T}(x_1|x_2,t_2,p(x_2,t_2))T(x_2,t_2|x_{0},t_0,p(x_0,t_0))}{\omega(p(x_2,t_2))}  \right)
\label{eq:the-equation}
\end{eqnarray}
As shorthand write  $w(x_2|x_1;t_1)= \tilde{T}(x_2|x_1,p(x_1,t_1))/\omega(p(x_1,t_1) = W(x_2|x_1;t_1)$, multiply both sides of
\cref{eq:the-equation} by $p(x_0,t_0)$, and integrate over $x_0$. Then we get the final equation
\begin{equation}
\frac{\mathrm{d} p(x_,t)}{\mathrm{d} t} = \sum_{y} \left[w(x|y,t) \Omega(p(y,t))-w(y|x,t) \Omega(p(x,t))\right]
\end{equation}
which is the non-linear master equation from the main text. Note that $w(x|y,t)$ cannot be obtained as the infinitesimal transition probability, but it is rather obtained as
\begin{equation}
w(x|y,t) = \lim_{\Delta t \rightarrow 0} \frac{W(x;t+\Delta t|y;t)- W(x;t|y;t)}{\Delta t} \, .
\end{equation}
Nevertheless, we still call $w$ as \emph{transition rate} since it plays similar role to the case of linear continuous-time Markov chain.

\section{Proof of theorem 1}
\begin{proof}

It will be useful to introduce the notation $w_{mn}\Omega(p_n) = J^\nu_{mn}$ and
$C_f = f'(\sum_m g(p_m))$.
The time derivative of the entropy is
\begin{eqnarray}
\dot{\mathbb{S}}(t) &=& C_f\sum_m g'(p_m) \dot{p}_m = C_f\sum_{m,n,\nu} g'(p_m)  (J^\nu_{mn} - J^\nu_{nm}) 
\label{eq:18}
\end{eqnarray}
We now divide this expression into a sum of two terms:
\begin{eqnarray}
\dot{\mathbb{S}}(t) =  C_f \sum_{m,n,\nu} (J^\nu_{mn} - J^\nu_{nm})\Phi_{mn}^\nu
+ C_f \sum_{m,n,\nu} (J^{\nu}_{mn} - J^\nu_{nm})
(g'(p_m)-\Phi_{mn}^\nu)
\end{eqnarray}
{ where $\Phi_{mn}^\nu$ is yet undetermined.}

{Next, we focus on the first term, identifying it as the entropy production rate. First, we rewrite it in the more suggestive form:
\begin{equation}
\dot{S}_i =  \frac{C_f}{2} \sum_{m,n,\nu} (J^\nu_{mn} - J^\nu_{nm})(\Phi_{mn}^\nu-\Phi_{nm}^\nu)
\end{equation}
According to Reqirement 3, $\dot{S}_i$ is non-negative and zero only if all currents vanish. First, the constant $C_f$ is positive because $f$ is increasing. Second, the non-negativity of the sum can be assured by considering non-negativity of each term. ** a little discussion**
From this, we have that $\Phi_{mn}^\nu < \Phi_{nm}^\nu$ if $J^\nu_{mn} < J^\nu_{nm}$ and vice versa.
Thus, we can conclude that $\Phi_{mn}^\nu$ is a function of $J^\nu_{mn}$, i.e., $\Phi_{mn}^\nu = \phi(J^\nu_{mn})$. Again, because we can relabel the probabilities, the function $\phi$ does not explicitly depend on $n,m$ or $\nu$.
}


Now we focus on the second term, rewriting it in the more suggestive form
\begin{equation}
    \frac{C_f}{2} \sum_{m,n,\nu} (J^{\nu}_{mn} - J^\nu_{nm})
[(g'(p_m)-\Phi_{mn}^\nu)-(g'(p_n)-\Phi_{nm}^\nu)]
\end{equation}
If the first term is the entropy production, then the second term is equal to entropy flow rate. Entropy flow is equal to
\begin{equation}
\dot{S}_e = \frac{1}{2} \sum_{mn\nu} (J^{\nu}_{mn} - J^\nu_{nm})\frac{ \epsilon_m-\epsilon_n}{T^\nu}.
\end{equation}
Write the difference of $\dot{S}_e$ and the second term as
\begin{eqnarray}
\frac{1}{2}  \sum_{\nu n m} (J_{mn}^\nu - J_{nm}^\nu)  \left\{C_f\left[(g'(p_m)-g'(p_n)) -  ({ \phi}(J^\nu_{nm})-{ \phi}(J^\nu_{mn}))\right] - \frac{\epsilon_n-\epsilon_m}{T^\nu}\right\}.\nonumber\\
\end{eqnarray}
This sum can be positive or negative, depending on $p_m$. Since we require that $\dot{S}_i$ must be non-negative, and vanishing only if the probability currents vanish, the difference must be zero, regardless of $p_m$. Since $(J_{mn}^\nu - J_{nm}^\nu)$ can be positive or negative depending on $p$, it means that the arguments in the curly brackets must be 1) independent of $p$ and 2) identically zero. Therefore we get that
\begin{equation}\label{eq:re}
   C_f\left[(g'(p_m)-g'(p_n)) -  ({\phi} (J^\nu_{nm})-{ \phi}(J^\nu_{mn}))\right]= \frac{\epsilon_n-\epsilon_m}{T^\nu}.
\end{equation}
The left-hand side is now depending on $p$, which should not be the case. This means that relation in square brackets does not depend on $p_m$, from which we immediately get
\begin{equation}\label{eq:req}
    { \phi}(J^\nu_{nm}) = j^\nu(w^\nu_{nm}) - g'(p_m)
\end{equation}
{ or more suggestively}
\begin{equation}
 J^\nu_{nm} = { \psi}(j^\nu(w^\nu_{nm}) - g'(p_m))
\end{equation}
where $J^\nu$ is an arbitrary function { and $\psi = \phi^{-1}$}. Since $J^\nu_{nm} = w_{nm} \Omega(p_m)$ we need to solve the following functional equation
\begin{equation}
{ \psi}(j^\nu(w^\nu_{nm}) - g'(p_m)) = w_{mn}^\nu \Omega(p_m)
\end{equation}
for unknown functions $\psi$ and $j^\nu$. This is a Cauchy multiplicative functional equation which determines $\psi$ and $j^\nu$ as $\psi(z) = \exp(z)$ and $j^\nu(z) = \log(z)$ (see \cite{Aczel96book}).  By plugging into the equation, we immediately obtain that
\begin{equation}
\Omega(p_m) = \exp(-g'(p_m))\, .
\end{equation}
By plugging the form of $\Omega$ into the local detailed balance, we immediately obtain that $\beta^\nu = \frac{1}{T^\nu}$. This result holds for any $g$. This concludes the proof.
\end{proof}

\section{Fokker Planck thermodynamics for nonlinear master equations}

Here we make a connection between the non-linear master and the corresponding Fokker-Planck equation on continuous spaces. Consider a general Master equation in the form Eq. (20). Consider a continuous state space, and rewrite the master equation (20) in the following form:
\begin{align}
\dot{p}(x,t) =
\int \mathrm{d}r \left[ w(x-r|r) \Omega(p(x-r;t)) - w(x|-r) \Omega(p(x;t)) \right]
\end{align}
Proceeding in the usual way, expand the first term in terms of $(x-r)$ around $r=0$.
Zeroth term cancels with the second term in the previous equation and by truncation after the term quadratic in $r$, we obtain that
\begin{eqnarray}
\dot{p}(x,t) &=& - \int \mathrm{d}r \, r \frac{\partial}{\partial x} [w(x|r) \Omega(p(x;t))] \nonumber\\&+& \frac{1}{2} \int_r \mathrm{d}r \, r^2 \frac{\partial^2}{\partial x^2} [w(x|r) \Omega(p(x;t))]\nonumber\\
&=& - \frac{\partial}{\partial x}[a_1(x) \Omega(p(x;t))] \nonumber\\&+& \frac{1}{2} \frac{\partial^2}{\partial x^2}[a_2(x) \Omega(p(x;t))]
\end{eqnarray}
where $a_n(x) = \int \mathrm{d}r \, r^n w(x|r)$. Thus, we end with the non-linear Fokker-Planck equation
\begin{equation}
\label{eq:fpe}
\dot{p}(x,t) = - \frac{\partial}{\partial x} J(p(x,t),x,t)
\end{equation}
where
\begin{equation}\label{eq:flow}
J = u(x,t) \Omega(p(x,t)) + D(x,t) \Omega(p(x,t)) \frac{\partial}{\partial x} g'(p(x,t))\, .
\end{equation}
This equation is called the \emph{Free energy nonlinear Fokker-Planck equation} \cite{Frank04book} due to its connection with thermodynamic quantities, as shown below. For simplicity, consider
the case where $f(x) = x$, i.e., where the entropy has the trace-class form
\begin{equation}
\mathbb{S}(t) = \int \mathrm{d} x \, g(p(x,t)).
\end{equation}
Evaluating, the time derivative of the entropy
\begin{eqnarray}
\dot{\mathbb{S}}(t) &=& \int \mathrm{d} x \,  g'(p(x,t)) \dot{p}(x,t) \nonumber\\&=& \int \mathrm{d} x \, J(p(x,t),x,t) \, \frac{\partial}{\partial x} g'(p(x,t))
\end{eqnarray}
The last expression was obtained from integration by parts. By plugging in \cref{eq:flow}, we get
\begin{equation}
\dot{\mathbb{S}}(t) =  \int \mathrm{d} x \, J(p,x,t) \left(\frac{J(p,x,t)}{D(x,t) \Omega(p(x,t))} -  \frac{u(x,t)}{D(x,t)} \right)\, .
\end{equation}
Here we can identify, similarly to the case of Shannon entropy \cite{Broeck10}, the entropy flow rate as
\begin{equation}
\dot{S}_e(t) = - \int \mathrm{d} x \, \frac{J(p(x,t),x,t) u(x,t)}{D(x,t)}
\end{equation}
and irreversible entropy production rate as
\begin{equation}
\dot{S}_i(t) = \int \mathrm{d} x \frac{J^2(p(x,t),x,t)}{D(x,t) \Omega(p(x,t))} \geq 0.
\end{equation}

\section{Derivation of the second law of thermodynamics on the trajectory level}

Let us consider a stochastic trajectory $n(\tau)$ with jumps in $t_j$ from $n_j^-$ to $n_j^+$. Let us calculate the time derivative of the stochastic entropy as
\begin{eqnarray}
\dot{s}(n(\tau)) = - \frac{\Omega'(p_{n(\tau)}(\tau))}{\Omega(p_{n(\tau)}(\tau))} \partial_\tau p_{n(\tau)}(\tau) \nonumber\\- \sum_j \delta(\tau-t_j) \ln \frac{\Omega(p_{n_j^+}(\tau))}{\Omega(p_{n_j^-}(\tau))}
\end{eqnarray}
which can be decomposed into entropy production rate
\begin{eqnarray}
\dot{s}_i(n(\tau)) =  - \frac{\Omega'(p_{n(\tau)}(\tau))}{\Omega(p_{n(\tau)}(\tau))} \partial_\tau p_{n(\tau)}(\tau) \nonumber\\ - \sum_j \delta(\tau-t_j) \ln \frac{w_{n_j^+,n_j^-}\Omega(p_{n_j^+}(\tau))}{w_{n_j^-,n_j^+} \Omega(p_{n_j^-}(\tau))}
\end{eqnarray}
and entropy flow rate
\begin{equation}\label{eq:slt}
\dot{s}_e(n(\tau)) =  - \sum_j \delta(\tau-t_j) \ln \frac{w_{n_j^+,n_j^-}}{w_{n_j^-,n_j^+}}
\end{equation}
Thus, we get that $\dot{s}(n(\tau)) = \dot{s}_i(n(\tau)) + \dot{s}_e(n(\tau))$. From the local detailed balance, we directly obtain that $\dot{s}_e(n(\tau)) = \sum_\nu \beta^\nu q^\nu(n(\tau))$.  The ensemble second law can be obtained by averaging Eq. \eqref{eq:slt} over all trajectories.
\section{Derivation of the detailed fluctuation theorem}

Here we focus on the log-ratio of probability distributions from the main text (Eq. (38)). The stochastic trajectory $n(\tau)$ can be taken as the limit of a discrete trajectory
$(n_k,t_k;n_{k-1},t_{k-1};,\dots,n_0,t_0)$. Express the log-ratio of multi-point probability distribution with help of Eq. \eqref{eq:nlmc}:
\begin{equation}
\log \frac{P(n_k,t_k;\dots,n_0,t_0)}{\tilde{P}(n_0,t_k;\dots,n_k,t_0)} = \log \frac{\prod_{j=0}^k W(n_j,t_j|n_{j-1},t_{j-1}) \Omega(p_{n_0}(t_0))}{\prod_{j=0}^k W(n_{j-1},t_{j}| n_j,t_{j-1}) \Omega(p_{n_k}(t_0))} = \log \Omega(p_i) - \log \Omega(p_f) + \sum_{j=0}^k \log \frac{W(n_j,t_j|n_{j-1},t_{j-1})}{W(n_{j-1},t_{j}| n_j,t_{j-1})}\, .
\end{equation}
The first term corresponds to the change of the trajectory entropy from the initial to the final state and the second term corresponds to change of entropy flow along the trajectory. Thus, the log-ratio is equal to the change of entropy production $\Delta s_i$ along the stochastic trajectory.

\section{Thermodynamic uncertainty relations and speed limits for non-linear systems}
Here we briefly show the exact connection to recent results regarding thermodynamic uncertainty relations and speed limits for the case of non-linear systems. We first mention the time-dissipation uncertainty relation \cite{Falasco20a}. By defining survival probabilities $p^s(t)$ as in \cite{Falasco20a}, we can derive an integral fluctuation theorem for $p^s(t)$ as a consequence of the detailed fluctuation theorem in the main text:
\begin{equation}
p^s(t) = \tilde{p}^s(t) \left\langle \exp e^{-\Delta s_i} \right\rangle_{\tilde{\textbf{s}}}
\end{equation}  
By defining the instantaneous rate $r(t):= - \frac{1}{p^s(t)} \frac{\mathrm{d} p^{s(t)}}{\mathrm{d} t}$, we recover the main result of \cite{Falasco20a}
\begin{equation}
\langle \dot{s}_i \rangle_{\tilde{\textbf{s}}}(t) \geq r(t) - \tilde{r}(t)
\end{equation} 
where $\dot{s}_i$ is the entropy production rate from the main text.

Second, we show that the results in \cite{Shiraishi18} can be straightforwardly generalized to the case of systems governed by the non-linear master equation. In order to do so, we introduce the dynamical activity of the system as
\begin{equation}
A(t) := \sum_{n \neq m} w_{mn}\Omega(p_n)
\end{equation}
and define the total entropy production of the system as $S_i(\tau) = \int_0^\tau \mathrm{d}t \dot{S}_i(t)$ where $\dot{S}_i(t)$ is defined in Eq. (13) of the main text. Let us also define $\langle A \rangle_\tau = \frac{1}{\tau} \int_0^\tau \mathrm{d} t A(t)$ and the variation distance as $L(p,p') := \sum_m |p_m-p'_m|$. Then by a straightforward calculation analogous to the proof in \cite{Shiraishi18} we obtain the speed limit in the following form:
\begin{equation}
\frac{L(p(0),p(\tau))^2}{2 S_i(\tau) \langle A \rangle_\tau} \leq \tau.
\end{equation}

%
%
%
%
%
\end{document}